\documentstyle[aps,10pt]{revtex}

\def \f{\frac}

\let \pp=\partial

\def \ni{\noindent}

\def \mn{\medskip
\ni}
\def \tr{\textrm}

\def \w{\wedge}
\newcommand{\Ref}[1]{(\ref{#1})}

\def \tl{\tilde}
\def \beq{\begin{equation}}
\def \eeq{\end{equation}}

\begin{document}

\title{\Large \bf Immirzi parameter in the Barrett-Crane model?}
\author{\bf Richard E. Livine} 
\address{Centre de Physique Th\'eorique - CNRS, Case 907, Luminy,
F-13288 Marseille, France \\
and \\
The Blackett Laboratory, Imperial College of Science, Technology
and Medecine, \\
South Kensington, London SW7 2BZ, UK}
\date{March 22, 2001}
\maketitle

\begin{abstract}
We study the generalised constrained BF theory described in
\cite{merced} in order to introduce the Immirzi parameter
in spin foam models. We show that the resulting spin foam
model is still based on simple representations and that
the generalised BF action is simply a deformation of the
Barrett-Crane model. The Immirzi parameter doesn't change
the representations used in the spin foam model,
so it doesn't affect the geometry of the model.
However we show how it may still appear as a factor in the area spectrum. 
\end{abstract}

\section{Introduction}

\ni
Spin foam models are an attempt to quantize general relativity
in a covariant way. They can be seen as a theory of evolving spin
networks \cite{mike&carlo}.
The most sucessful model is the Barrett-Crane model
\cite{BC} in Euclidian general relativity, which has shown to have some
very interesting finiteness properties \cite{alej2}.
There is also a similar model
in Lorentzian general relativity \cite{BC,alej}. They are based on the
simple representations of the symmetry algebra $so(4)$ and of $so(3,1)$.
There are two ways of deriving the Barrett-Crane model as a theory of
gravity (or quantum geometry).
First, there is a
geometrical method, which relies on a
simplicial triangulation of the space-time 4-manifold and uses the
isomorphism between the set of bivectors and the rotation Lie algebra
($so(4)$ or $so(3,1)$)
to associate to each triangle of the decomposition a (simple)
representation \cite{BC,baez}. 
Then, there is the approach by discretisation of a constrained
BF theory \cite{laurent1,daniele}. The constraint restricts
us to the use of simple representations again.

\mn
Recently, a generalised constrained action has been proposed by
Capovilla, Montesinos, Prieto and Rojas \cite{merced}
in order to introduce an Immirzi parameter as in the canonical approach
(loop quantum gravity) \cite{immirzi,tom&carlo}.
It seemed, as suggested in \cite{merced}, that discretising that
new BF action would lead us to introduce non-simple representations
in spin foam models. Then, checking if those models be consistent or not,
it could lead to some constraints on the possible values of
the Immirzi parameter,
or else we would have an arbitrary parameter in spin foam
models as in loop quantum gravity.

\mn
Here, I will show that one might, instead, reconsider
the way the bivector $B$ field is associated to
generators of the symmetry algebra. Instead of associating it as
usual with the canonical generators of $so(4)$ (or $so(3,1)$), we are
lead to make a change of basis which allow us to keep the
simple representations.
I then argue that this could have been seen
directly by a change of variables on the initial action.
The Immirzi parameter then doesn't have any effect on
the representations used in the spin foam model, so that
we are still studying the same Barrett-Crane model.
I conclude with a calculation of the area spectrum, by trying to
apply canonical methods to the BF action,
and show how the Immirzi parameter can still
introduce a factor between the area and the non-zero Casimir operator
of $so(4)$ in the Euclidian case or $so(3,1)$ in the Lorentzian case.

\section{Euclidian BF theory with Immirzi parameter}

\label{debut}

\ni
I start by recalling  the  action proposed in \cite{merced} for the
constrained BF theory for Euclidian general relativity:

\begin{equation}
S=\int B^{IJ}\w F_{IJ} -\f{1}{2}\phi_{IJKL}B^{IJ}\w B^{KL}
+\mu H
\label{action}
\end{equation}

\ni
where $H=a_1\phi_{IJ}\,^{IJ}+a_2\phi_{IJKL}\epsilon^{IJKL}$.
$\phi$ and $\mu$ are Lagrange multipliers and $\phi$ has
the symmetries $\phi_{IJKL}=-\phi_{JIKL}=-\phi_{IJLK}=\phi_{KLIJ}$.
$I,J,K,L$ are internal indices. The $\wedge$ product is defined
on the space-time indices and the $*$ operator acts on internal indices:
$*B_{IJ}=1/2\epsilon_{IJKL}B^{KL}$ and $*^2=1$.

\ni
The same action can also be used for Lorentzian
general relativity with the same
definitions. The difference is that $*^2=-1$ in the Lorentzian case, so that
all the equations will be the same up to a sign.

\mn
The constraints on the field B are then:

\beq
B^{IJ}\w B^{KL} =\f{1}{6} (B^{MN}\w B_{MN}) \eta^{[I |K|} \eta^{J]L}
+\f{1}{12}(B^{MN}\w *B_{MN}) \epsilon^{IJKL}
\label{whole}
\eeq

\begin{equation}
2a_2 B^{IJ}\w B_{IJ} -a_1 B^{IJ}\w *B_{IJ}=0
\label{simple}
\end{equation}

\ni
Solving these constraints for the field B \cite{prieto}, we get:

\begin{equation}
B^{IJ}=\alpha *(e^I \w e^J) + \beta e^I \w e^J
\label{B}
\end{equation}

\ni
where

\beq
\f{a_2}{a_1}=\f{\alpha^2+\beta^2}{4\alpha\beta}
\label{a1a2}
\eeq

\ni
Replacing this expression in the original action, we get
an action where the general relativity action is coupled to
the ``trivial'' sector:

\beq
S=\alpha \int  *(e^I \w e^J)\w F_{IJ}  +
\beta \int e^I \w e^J \w F_{IJ}
\eeq

\ni
Let's note that the equations of motion ignore the ``trivial'' part and
still gives the Einstein equations.

\mn
We can have a look at the particular cases for which we take only one
of the two terms in the constraint $H$.
For $a_2=0$, we get $\alpha=\beta=0$, it is a
``empty'' theory. In fact, only degenerate tetrads are going to contribute
to the action.
For $a_1=0$, we have $\alpha=0$ or $\beta=0$, which means
that we have
two distinct sectors of solutions: the general relativity one
$*(e\wedge e)$ and the ``trivial'' one $e\wedge e$. More
precisely, we have four sectors \cite{so4}: $\alpha=\pm 1,\beta=0$ and
$\alpha=0, \beta=\pm 1$.

\mn
Following \cite{holst},
the Immirzi parameter is $\gamma=\alpha/\beta$.
The equation relating $a_1,a_2$ and $\gamma$ is:

\beq
\f{a_2}{a_1}=\f{1}{4}\left(\gamma +\f{1}{\gamma}\right)
\eeq

\ni
There are two solutions to this equation, inverse of each other.
We have two sectors in our theory with Immirzi parameter $\gamma$ or
$1/\gamma$: there is a symmetry exchanging $\alpha$ and $\beta$.
I will call that symmetry the $*$-symmetry since the resulting $B$ field
get changed into its image by the Hodge operator $*$ under the exchange
of $\alpha$ and $\beta$.

\mn
To introduce these modifications into the spin foam,
we rewrite \Ref{simple} in an equivalent form:

\begin{equation}
2a_2\f{1}{2}\epsilon_{IJKL}B^{IJ}_{ab}B^{KLab}
-a_1B^{IJ}_{ab}B^{ab}_{IJ}=0
\label{simple2}
\end{equation}

\ni
Indeed, we can get to \Ref{simple2} from \Ref{simple} by changing the
role of the internal indices and space-time indices. What makes the two
equations equivalent is that they have the same set of solutions, given by
\Ref{B} (see in appendix for more details)
We could also deal directly with
the condition \Ref{simple} as shown by Reisenberger \cite{mike,reis} and
by De Pietri and Freidel \cite{so4} in the case $a_1=0$, and
it would lead to (a modification of) the Reisenberger spin foam model,
which is a different model than the Barrett-Crane one.

\ni
Then, we can translate
the above constraint  on the field B
into a constraint on the two Casimirs of $so(4)$,
using the equivalence between
this field and the generators of the
$so(4)\approx su(2)_L\oplus su(2)_R$
algebra discussed in the framework of spin foam models \cite{BC}.
More precisely, replacing $B^{IJ}_{ab}$ by the canonical
generators $J^{IJ}$ of $so(4)$, we get:

\beq
\begin{array}{ccc}
B^{IJ}_{ab}B^{ab}_{IJ} &
\rightarrow & J^{IJ}J_{IJ}= 2((J^{01})^2+\dots)=2C_1\\
\f{1}{2}\epsilon_{IJKL}B^{IJ}_{ab}B^{KLab} &
\rightarrow & \f{1}{2}\epsilon_{IJKL}J^{IJ}J^{KL}
=4(J^{01}J^{23}+\dots)=2C_2
\end{array}
\eeq

\ni
Using the above correspondence, we get to a modified simplicity constraint:

\begin{equation}
2a_2 C_2 - a_1 C_1 =0
\label{mixed}
\end{equation}

\ni
or equivalently:

$$
2\alpha\beta C_1 = (\alpha^2+\beta^2)C_2
$$

\ni
Then we could conclude as stated in \cite{merced}
that we should use non-simple representations
in our spin foam model. This conclusion is problematic
in the Euclidian case where $C_1$ and $C_2$ are discrete.
More precisely, in the Euclidian case we are studying,
an equation $xC_1=yC_2$ has an infinite number of solutions
only in 3 cases.
If $x=y$, we get the representations of $su(2)_L$
($j^-=0$). If $x=-y$, we use the representations of $su(2)_R$.
And finally $x=0$ gives us the simple representations
of $so(4)$. In the other cases, we have in general no solutions, or
sometimes only one solution (one representation $(j^+,j^-)$), which
would give quite an ill-defined spin foam model.

\ni
Apart from those problems,
it is hard to understand why we should give up the
``simple representations'' which are understood to be the
simplest and most natural representations of $so(N)$ \cite{laurent2}.

\ni
In the next section, I describe an alternative:
we can identify
the $B$ field to some generators of the $so(4)$ algebras
in such a way to that we keep
the simple representations. This correspondence
is at the heart of the ``spin foam quantization'' because it is through
it that we translate the simplicity (and intersection) constraint
to the quantum level.

\section{Modifying the correspondence $B\rightarrow J$}

\label{corres}

\ni
We could directly replace $B^{IJ}$ by the
canonical generator $J^{IJ}$ of $so(4)$. This would give us the
mixed simplicity condition \Ref{mixed}.
On the other hand,
looking at \Ref{B}, we can associate the following generators
to the field $B$:

\begin{equation}
\tl{J}^{IJ}=\alpha J^{IJ}+\beta *J^{IJ}
\label{modif}
\end{equation}

\ni
This is simply a change of basis. It is still the same
algebra $so(4)$ we are dealing with, in the same chosen representation
we would be studying.
I will call a choice of basis ``consistent'' (with the
Barrett-Crane simplicity constraint) if it is
a change of basis which leads to the simple representations
after translating the simplicity constraint \Ref{simple2}
to the Lie algebra generators.
\Ref{modif} is a consistent choice as I show below. It
corresponds to associating $J^{IJ}$
to $*(e^I\wedge e^J)$ (or to $e^I\wedge e^J$)
instead of associating it directly to the bivector field $B^{IJ}$.

\ni
There are other consistent
choices of $\tl{J}^{IJ}$. First, we can exchange $\alpha$ and $\beta$
in the above formula. This doesn't seem to affect anything in the theory.
We could also rotate (by an $so(4)$ rotation) $\tl{J}^{IJ}$. This wouldn't
change the Casimir, so it wouldn't affect the discussion below. And also,
we could rescale $\tl{J}^{IJ}$ by a factor $\lambda$, this corresponds
to a rescaling of the frame field $e^I_a$. It rescales the Casimirs by
a factor $\lambda^2$ and doesn't seem to affect any result, except maybe
the spectrum of the area as shown in section \ref{area}.
There are also other
changes of basis which don't affect any of the following result like
changing the signs of the generators $J^{0i}$. but, we aren't going
to study these here.

\ni
Taking a non-consistent choice of basis would lead to a mixed simplicity
condition $xC_1=yC_2$ which doesn't give any self-consistent spin foam
model in general. Moreover the choice \Ref{modif} is natural as we
replace the geometrical meaningful $*(e^I\wedge e^J)$ (general relativity
sector) by the canonical generators  $J^{IJ}$.

\mn
The Casimir corresponding to $B^{IJ}\wedge B_{IJ}$ is then

\beq
\tl{C}_2=\f{1}{4}\epsilon_{IJKL}\tl{J}^{IJ}\tl{J}^{KL}
=2\alpha \beta C_1 +(\alpha^2+\beta^2)C_2
\eeq

\ni
where $C_1$ and $C_2$ are the usual Casimirs associated to the generators
$J^{IJ}$:

$$
C_1=\f{1}{2}\left((J_i^+)^2 +(J_i^-)^2\right)
=\left(j^+(j^+ +1)+j^-(j^- +1)\right)
$$

$$
C_2=\f{1}{2}\left((J_i^+)^2 -(J_i^-)^2\right)
=\left(j^+(j^+ +1)-j^-(j^- +1)\right)
$$

\ni
Also the Casimir associated to $B^{IJ}\wedge *B_{IJ}$ becomes

\beq
\tl{C}_1=\tl{J}^{IJ}\tl{J}_{IJ}
=(\alpha^2+\beta^2)C_1 + 2\alpha \beta C_2
\eeq

\ni
Now writing the simplicity condition \Ref{simple}, we get the equation:

\beq
2\alpha\beta \tl{C}_1=(\alpha^2+\beta^2)\tl{C}_2 \Rightarrow
(\alpha^2-\beta^2)^2C_2=0
\eeq

\ni
And in the case $\alpha^2\ne \beta^2$ ($\gamma^2\ne 1$),
we find back the usual
simplicity condition $C_2=0$ and the simple representations
given by $j^+=j^-$. And the whole modification of the inital action
is absorbed by a suitable redefinition of the correspondence between
the field $B$ and the generators of the Lie algebra.
In the degenerate case $\gamma=\pm 1$
(Barbero's choice for canonical gravity),
we have to see whether we want to impose directly by hand
the condition $C_2=0$ or maybe impose something else as discussed
in section \ref{phirotate}.
The degeneracy comes from the fact that the $*$-symmetry
$(\alpha,\beta)\rightarrow(\beta,\alpha)$ is
exactly realised when $\alpha=\beta$.

\mn
We can interpret the Casimir $C_1$ of a representation as
the squared area associated to it.
A more rigourous proof of this is done in section \ref{area}
through a canonical treatment. But here, in our covariant
framework, there is no reason to introduce an Immirzi parameter
in the definition of the area (we can
rescale it as wished) and
we can define it in terms of Casimirs $\tl{C}$:

\beq
{\cal A}^2=C_1=\f{1}{(\alpha^2-\beta^2)^2}
\left((\alpha^2+\beta^2)\tl{C}_1-2\alpha\beta\tl{C}_2\right)
\eeq

\ni
This definition allows us to relate the area  with the correlation
of the $B$ field that we could calculate using the spin foam model.
It could be interesting to check whether calculating those
correlations directly (as in \cite{laurent1})
would yield the same result as above (the Casimir).
In fact, the two results are likely to be different.
Already in two dimensions and three dimensions (Euclidian), calculating
the length associated to a edge labelled by a $su(2)$ representation $j$
as a correlation of the discretised BF theory
gives us $j+1/2$ instead of the expected $\sqrt{j(j+1)}$ \cite{laurent1}.


\mn
Let's point out that
in the Euclidian case, we could think to simply associate
the self-dual part of the B field with the self-dual
generators and the same for the anti-self dual part.
That works up to a normalisation.
More precisely,

$$
B=\alpha *E +\beta E \Rightarrow B=B_+ +B_-
\tr{ with }
\left\{
\begin{array}{ccc}
B_+&=&\f{\alpha+\beta}{2}(E+*E) \\
B_-&=&\f{\alpha-\beta}{2}(E-*E)
\end{array}
\right. 
$$

\ni
The condition \Ref{simple2} then writes
$(\alpha-\beta)^2(B_+)^2=(\alpha+\beta)^2(B_-)^2$.
And we realise that we must normalise
a correspondence $B_\pm\rightarrow J_\pm$ by
the factors $(\alpha-\beta)$ and $(\alpha+\beta)$. This causes
some problem when $\alpha=\pm \beta$ as in the beginning
of the section.
However, when $\alpha\ne\pm \beta$, we find directly the
condition $(J_+)^2=(J_-)^2$ which is the simplicity condition
$C_2=0$.

\section{Link with the geometrical picture}
\label{geompic}

\ni
The geometrical picture consists of a simplicial triangulation
of the 4-manifold (into 4-simplices). We then associate to
each triangle (2-simplex) a bivector $B$. The Barrett-Crane model
\cite{BC} associates to a triangle defined by its edges $e,f,g$
the bivector $B=e\wedge f = f \wedge g = g \wedge e$. This
is a simple bivector. This can be linked up to
a discretisation of the constrained BF action for $a_1=0$.
In the mixed case, the $B$ field is not simple anymore,
it satisfies the relation \Ref{simple} and is given by
the equation \Ref{B}. Discretising it, we can translate the
constraint \Ref{simple} by a constraint on the bivector $B$:

\beq
2\alpha\beta B\wedge B =(\alpha^2+\beta^2)B\wedge *B
\eeq

\ni
This leads to the solution

\begin{equation}
B=\alpha E + \beta *E
\label{bivect}
\end{equation}

\ni
where $E$ is a simple bivector ($*E$ is then also a simple bivector).
This solution corresponds to a discretisation of the classical solution
\Ref{B}.

\mn
For two triangles which
share an edge, we must impose an intersection constraint stating that
$B=B_1+B_2$ should still be of the mixed simple form \Ref{bivect}.
It is easy to check
that the corresponding condition can be written:

\beq
2\alpha\beta B_1\wedge B_2 =(\alpha^2+\beta^2)B_1\wedge *B_2
\eeq

\ni
and that in that case, the simple bivector $E$ associated to $B$
is simply $E_1+E_2$. This shows us that although we have modified
the field (or bivector) $B$, everything is still working as if we
were using simple bivectors 
(the field or bivector $E=(\alpha B -\beta *B)/(\alpha^2-\beta^2)$).
The simple bivectors (and therefore simple representations)
give the whole structure of the theory.
We can also see that directly on the action \Ref{action} as
shown in the following section.

\section{From mixed variables to pure simple variables}

\label{phirotate}

\ni
Starting from the initial action \Ref{action}, there is a change of
variables which makes it obvious that the constraint on the field B
is still the pure simplicity constraint.
We start by finding $\alpha$ and $\beta$ verifying \Ref{a1a2} for some
coefficient $\lambda$:

$$
\begin{array}{cc}
a_1=&\lambda (4\alpha \beta) \\
a_2=&\lambda (\alpha^2+\beta^2)\\
\end{array}
$$

\ni
Then, we can do the change of variables
$(B,\phi)\rightarrow (E,\tl{\phi})$:

\begin{equation}
\left\{
\begin{array}{ccc}
B^{IJ}& =&\alpha E^{IJ}+\beta *E^{IJ} \\
\tl{\phi}_{IJKL}&=&
(\alpha+\beta\f{1}{2}\epsilon_{IJ}\,^{AB})\phi_{ABCD}
(\alpha+\beta\f{1}{2}\epsilon_{KL}\,^{CD})
\end{array}
\right.
\label{change}
\end{equation}

\ni
This change of variables is well-defined and invertible only if
$\alpha^2\ne\beta^2$. This excludes three possibilities:
$a_2=0$ and $2a_2=\pm a_1$ which we will discuss below.
When the change is well-defined, the action \Ref{action} becomes:

\begin{equation}
S=
\f{1}{|\alpha^2-\beta^2|^3}
\int (\alpha E^{IJ}+\beta *E^{IJ})\wedge F_{IJ}
-\f{1}{2}\tl{\phi}_{IJKL}E^{IJ}\wedge E^{KL}
+\mu\lambda\epsilon^{IJKL}\tl{\phi}_{IJKL}
\label{after}
\end{equation}

\ni
Looking at \Ref{after}, we notice that the simplicity constraint on
the bivector field $E^{IJ}$ is the pure simplicity constraint.
This shows us how to get to the correspondence \Ref{modif}
which lead from the mixed simplicity condition
to the pure simplicity condition.

\mn
The case $a_2=0$ leads to $\alpha=\beta=0$ and the set of solutions
for the field $B$ contains only degenerate tetrads. The cases
$2a_2=\pm a_1$ are more interesting. They correspond to
the self-dual and the anti-self dual cases. As we have seen earlier,
they do not lead directly to the simplicity constraint
$C_2=0$ after the change of basis of the generators $J^{IJ}$
\Ref{modif}. So we have many possibilities. First, we could impose by
hand the pure simplicity constraint and say the model is
equivalent once more to the Barrett-Crane model (at an infinite coefficient).
Or we could forget the change of basis. In fact, as it corresponds
to the change of variables \Ref{change}, it might be ill-defined.
Keeping $B^{IJ}\rightarrow J^{IJ}$ where $J^{IJ}$ is the canonical basis,
we get modified simplicity constraints: $C_1=C_2$ for $2a_2=a_1$
and $C_1+C_2=0$ for $2a_2=-a_1$. The first case leads to $j^-=0$ and
the second to $j^+=0$. So we are reduced from $so(4)$ to
$su(2)_L$ or to $su(2)_R$.
What theory do we then get? 
If we want to build a spin foam model
similar to the Barrett-Crane model in 4 dimensions but
based on $SU(2)_L$ only (or $SU(2)_R$ only),
we would naturally be lead to the $\{15j\}$ and to
a topological theory corresponding to $SU(2)$
BF theory.
However, it can not be the topological $SU(2)$ BF theory since although
we are reduced to a $su(2)$ gauge symmetry, we
still have the constraint that the $B$ field should be derived
from a tetrad field, which is not implemented in the topological
theory.
More precisely, in the case we are reduced to the representations $j^+$
and therefore to the gauge group $SU(2)_L$, the $SU(2)_R$ degrees of freedom
of the connection decouple and we can forget them. We then get
the left-handed (self-dual) action for Euclidian gravity.
There exist a corresponding spin foam model:
it is the {\it Reisenberger
model for left-handed euclidian gravity} \cite{mike}.
We derive it discretising
the constraint \Ref{simple} and not its equivalent form \Ref{simple2}.
But this equivalence works only when $2a_2\ne \pm a_1$ (see appendix).
So, perhaps, although the Reisenberger model
is well-defined in both the full case \cite{reis,so4}
and the self-dual case \cite{mike,laurent1},
the Barrett-Crane approach is ill-defined in the self-dual case
and need to be regularisd by a choice
$\alpha$ slightly different than $\pm\beta$.
It is important to understand what happens precisely
for these degenerate cases
correspond classically to the self-dual (and anti-self-dual)
connection action which is usually used in the canonical treatment of
Euclidian general relativity.

\section{Lorentzian case}

\label{lorentz}

\ni
In the Lorentzian case, starting from the same action \Ref{action},
we can write the same equations as in the Euclidian case.
We are lead to the same conclusion that the representations we should use
are the simple ones. More precisely, there is a difference of sign in some
equations due to $*^2=-1$ instead of $*^2=1$. First, the equation
defining $\alpha$ and $\beta$ is:

\beq
\f{a_2}{a_1}=\f{\alpha^2-\beta^2}{4\alpha\beta}
\eeq

\ni
$a_2=0$ now corresonds to the case $\alpha^2=\beta^2$
while $a_1=0$ still corresponds to $\alpha=0$ or $\beta=0$.
The $*$-symmetry is now $(\alpha,\beta)\rightarrow(\beta,-\alpha)$
and $\gamma \rightarrow -1/\gamma$.

\ni
The simplicity constraint becomes:

\beq
2a_2 B^{IJ}\w B_{IJ} +a_1 B^{IJ}\w *B_{IJ}=0
\eeq

\ni
It is equivalent (same set of solutions) to the equation:

\beq
2a_2\f{1}{2}\epsilon_{IJKL}B^{IJ}_{ab}B^{KLab}
-a_1B^{IJ}_{ab}B^{ab}_{IJ}=0
\eeq

\ni
which can be translated to the Lie algebras generators:

\beq
2a_2\tl{C}_2-a_1\tl{C}_1=0 \Rightarrow
(\alpha^2-\beta^2)\tl{C}_2 = 2\alpha\beta\tl{C}_1
\eeq

\ni
It is the same modified simplicity condition as in the Euclidian case
expressed in terms of $a_1,a_2$.
We modify a sign in the change of basis:

$$
B^{IJ}\rightarrow\tl{J}^{IJ}=\alpha J^{IJ}-\beta *J^{IJ}
$$

\ni
The minus sign is normal since if we associate $J^{IJ}$ to
$*(e^I\w e^J)$ then we must
associate $-*J^{IJ}$ to $e^I\w e^J=-*(*(e^I\w e^J))$.

\ni
Then the modified Casimirs are:

\beq
\begin{array}{ccc}
\tl{C}_1 &
=& (\alpha^2-\beta^2)C_1 - 2\alpha \beta C_2 \\
\tl{C}_2 &
=& +2\alpha \beta C_1 +(\alpha^2-\beta^2)C_2
\end{array}
\eeq

\ni
The two Casimirs are \cite{BC,alej}, where $j$ is a positive half-integer
bigger than 1 and $\rho$ a positive real number:

$$
\begin{array}{ccc}
C_1&=&j^2-\rho^2-1 \\
C_2&=&2j\rho
\end{array}
$$

\mn
The modified simplicity condition then reads
$(\alpha^2+\beta^2)^2C_2=0$.
Since $\alpha^2+\beta^2\ne0$ in all real cases, 
we get to the pure simplicity condition $C_2=0$.
We have two series of simple representations: a discrete series given 
by $\rho=0$ and a continuous series defined by $j=0$.
We can note that the degenerate case
$\alpha^2+\beta^2=0$
would be the Ashtekar choice of
variables with $\alpha=1$, $\beta=i$ and $\gamma=\pm i$.
This choice of $\gamma$ makes the $*$-symmetry obvious in the
Lorentzian case. The Casimirs are then:

$$
\begin{array}{ccc}
\tl{C}_1 &
=& 2(C_1 + i C_2) \\
\tl{C}_2 &
=& -2i( C_1 +i C_2)
\end{array}
$$

\ni
We can do the same change of variables as in the Euclidian case.
In contrast with the Euclidian case,
{\it everything is always well-defined}. More precisely,
we need $\alpha$ and $\beta$ such that

$$
\begin{array}{cc}
a_1=&\lambda (4\alpha \beta) \\
a_2=&\lambda (\alpha^2-\beta^2)\\
\end{array}
$$

\ni
Then the change of variable
\Ref{change} is invertible if $\alpha^2+\beta^2\ne0$, which is
the case for any real $a_1,a_2$ coefficients. So we don't have any ambiguous
cases as in the Euclidian case. All the cases lead to the Lorentzian
Barrett-Crane model.

\section{Calculation of the area spectrum for $so(4)$ and
$so(3,1)$ spin networks}

\label{area}

\ni
It is interesting to calculate the area spectrum of the theory
to see if we find a factor corresponding to the Immirzi parameter
as in the case of loop quantum gravity \cite{tom&carlo}.
We could calculate the
the area expectation values from the spin foam action
as in \cite{laurent2}.
However, I choose a more
direct approach to compute the area spectrum for
this constrained BF theory.
I will use the same tools as
when calculating the area spectrum in the canonical theory
\cite{lee&carlo} and try to adapt it to use it on
$so(4)$ and $so(3,1)$ spin networks.
More precisely,
I will follow the canonical approach as described in
\cite{primer}. Nevertheless,
I will assume a natural quantification for the connections
and their conjugate momenta, check whether it gives us consistent
results and derive from it a Poisson bracket
that one should be able to find directly from the action.

\mn
We study the action:

$$
\begin{array}{ccccc}
S(A,e) &= &
\alpha \int  *(e^I \w e^J)\w F_{IJ}  &+&
\beta \int e^I \w e^J \w F_{IJ} \\
&= &
\alpha\int \epsilon^{abcd}\f{1}{2}\epsilon_{IJKL}
e_a^I e_b^J F_{cd}^{KL}
&+ & \beta\int \epsilon^{abcd}e_{aI} e_{bJ} F_{cd}^{IJ} \\
\end{array}
$$

\ni
The metric is given by $g_{\mu\nu}=e^I_\mu e_{I\nu}$.

\ni
We are working with a real $so(4)$ connection $A$, its
strength field $F$ and a real tetrad $e$.
The conjugate momentum of the variable $A_c^{IJ}$ is:

$$
\Pi^c_{IJ}=\epsilon^{abc}(
\alpha \f{1}{2}\epsilon_{IJKL}e_a^K e_b^L
+\beta e_{a[I} e_{bJ]}) \quad a,b,c=1,2,3
$$

\ni
In a first calculation, we are going to ignore the (simplicity)
constraints linking those conjugate momenta.

\ni
We are doing a hamiltonian treatment choosing $a=0$ as our time
direction and ${\cal M}$ parametrised by $x_i,\,i=1,2,3$ as our
initial hypersurface.
As we are interested by spin foams whose surfaces are labelled by
$so(4)$ representations, we are going to study
$so(4)$ spin networks.
More precisely,
the ``observables'' we consider are the holonomies of the connection
$A=A^{IJ}J_{IJ}$ so that our spin networks are graphs imbedded in 
${\cal M}$ and labelled by principal unitary
irreducible representations of
$so(4)$. The possible representations are restricted by the (modified)
simplicity condition discussed earlier.

\mn
We are considering
infinitesimal 2\_surface $\Sigma$ with the coordinate system
$\vec{\sigma}=(\sigma^1,\sigma^2)$ and we want to compute its area.
We will only consider  spin networks which intersect that
surface just once and never at a node.

\ni
$\Pi$ is the conjugate momentum of the connection $A$.
If we cheat and ignore the constraints,
the corresponding operator acting on the spin networks is given by: 

\begin{equation}
\Pi^a_{IJ}\rightarrow
-i\hbar \f{\delta}{\delta A^{IJ}_a} \approx \hbar J_{IJ}
\label{wrong}
\end{equation}

\ni
To get the area, we must integrate the induced 2\_metric on
$\Sigma$.
We introduce the following operator:

$$
{\cal B}(\Sigma)=
\sqrt{{\cal B}^{IJ}(\Sigma){\cal B}_{IJ}(\Sigma)}
$$

\ni
where

$$
{\cal B}_{IJ}(\Sigma)=\int_\Sigma d\sigma^1 d\sigma^2
\epsilon_{abc}
\f{\pp x^a(\vec{\sigma})}{\pp \sigma^1}
\f{\pp x^b(\vec{\sigma})}{\pp \sigma^2}
\Pi^c_{IJ}
$$

\ni
This operator is closely related to the area. To show this, let's
choose a special parametrisation of $\Sigma$.
Let's choose $x^3=0$ on $\Sigma$, $x^{1,2}=\sigma^{1,2}$. Then the
quantity ${\cal B}$ becomes:

$$
{\cal B}(\Sigma)=
\int_\Sigma d\sigma^1 d\sigma^2
\sqrt{
\epsilon_{abc}
\f{\pp x^a(\vec{\sigma})}{\pp \sigma^1}
\f{\pp x^b(\vec{\sigma})}{\pp \sigma^2}
\Pi^{cIJ}
\epsilon_{def}
\f{\pp x^d(\vec{\sigma})}{\pp \sigma^1}
\f{\pp x^e(\vec{\sigma})}{\pp \sigma^2}
\Pi^f_{IJ}
}
$$

$$
{\cal B}(\Sigma)=
\int_\Sigma d\sigma^1 d\sigma^2
\sqrt{
\Pi^{3IJ}\Pi^3_{IJ}
}
$$

\ni
Looking up at the definition of $\Pi$ and of $g_{\mu\nu}$, we get:

\beq
{\cal B}(\Sigma)=
\int_\Sigma d\sigma^1 d\sigma^2
\sqrt{(\beta^2+\alpha^2)
(g_{11}g_{22}-g_{12}g_{12})
}
=
\int_\Sigma d\sigma^1 d\sigma^2
\sqrt{(\beta^2+\alpha^2)
\tr{det}\,^2 g_{(\Sigma)}
}
\eeq

\ni
Promoting $\Pi$ to an operator and we get an operator ${\cal B}$
acting on $so(4)$ spin networks and whose values
are the area to a factor $\sqrt{(\beta^2+\alpha^2)}$.
Replacing the variable $\Pi$ by the generators of the $so(4)$ algebra
as in \cite{primer}, we find that the eigenvalues
of the operator ${\cal B}$  are given by the square root of
the Casimir $C_1=J^{IJ}J_{IJ}$ (without any extra sign) times $\hbar$.

\mn
To sum up the results, we get that the squared area is given by:

\beq
{\cal A}^2=
\f{1}{\beta^2+\alpha^2} C_1
\eeq

\ni
We could also decide to use the operator ${\cal C}$ defined as follows:

$$
{\cal C}(\Sigma)=
\sqrt{{\cal B}^{IJ}(\Sigma)\, *{\cal B}_{IJ}(\Sigma)}
$$

\ni
Following the same procedure as in the cas of the ${\cal B}$ operator,
we find:

\beq
{\cal C}(\Sigma)=
\int_\Sigma d\sigma^1 d\sigma^2
\sqrt{(2\alpha\beta)
\tr{det}\,^2 g_{(\Sigma)}
}
\eeq

\beq
{\cal A}^2=
\f{1}{2\alpha\beta} C_2
\eeq

\ni
So if we want the results to be consistent (${\cal A}^2={\cal A}^2$),
we must impose:

$$
2a_2 C_2 =a_1 C_1
$$

\ni
We see that assuming the canonical bracket between the field $B$ and the
connection $A$, we find the simplicity condition. But we have
showed earlier that this wasn't the ``right'' condition:
we need to use a new correspondence between $B$ and the Lie algebra,
changing the Casimirs $C_1,C_2$ into the new Casimirs $\tl{C}_1,\tl{C}_2$.
This means that we must change the assumed Poisson bracket between
$B$ and $A$ so that:

\begin{equation}
\Pi^a_{IJ}\rightarrow
-i\hbar \left(
\alpha\f{\delta}{\delta A^{IJ}_a}+\beta\f{\delta}{\delta *A^{IJ}_a}
\right)
\approx \hbar \tl{J}_{IJ}
\label{quant1}
\end{equation}

\ni
We then get the right simplicity condition 

$$
2a_2 \tl{C}_2 =a_1 \tl{C}_1
$$

\ni
and, as soon as this condition is fulfilled, the area is 

\beq
{\cal A}^2=
\f{1}{\beta^2+\alpha^2} \tl{C}_1
=\f{1}{2\alpha\beta} \tl{C}_2
=C_1
\eeq

\ni
And we find no Immirzi parameter. Still, there were different
consistent choices for $\tl{J}^{IJ}$. First, we could exchange
$\alpha$ and $\beta$:

\begin{equation}
\Pi^a_{IJ}\rightarrow
-i\hbar \left(
\beta\f{\delta}{\delta A^{IJ}_a}+\alpha\f{\delta}{\delta *A^{IJ}_a}
\right)
\label{quant2}
\end{equation}

\ni
That doesn't change anything. On another hand, we could rescale the
relations \Ref{quant1} and \Ref{quant2}
by a factor $\lambda$ and that would rescale the area (square)
by a factor $\lambda$ (square).

\mn
More precisely, to make the link
with the canonical formalism \cite{holst}, we can chose $\alpha=1$, and
then the Immirzi parameter is $\gamma=1/\beta$. Normalising
\Ref{quant1} to have a first term
$\gamma{\delta}/{\delta A^{IJ}_a}$,
we need a factor $\lambda=1/\beta=\gamma$. Then in that case,
we find the ``usual'' area spectrum:

\begin{equation}
\Pi^a_{IJ}\rightarrow
-i\hbar \left(
\f{1}{\beta}\f{\delta}{\delta A^{IJ}_a}+\f{\delta}{\delta *A^{IJ}_a}
\right)
\label{quant3}
\end{equation}

\beq
{\cal A}^2=\f{1}{\beta^2} C_1 = \gamma^2 C_1 \qquad
{\cal A}=\gamma \sqrt{C_1}=
\gamma\sqrt{2j^+(j^++1)}
\eeq

\ni
First, the choice \Ref{quant3} is consistent with the (classical)
simplicity constraint \Ref{simple} (replacing $B$ by $\Pi$) as soon
as we choose the simple representations $C_2=0$ in the holonomies
which are the state space of our quantum theory.
It works because the simple representations automatically verify
$2\alpha\beta\tl{C}_1=(\alpha^2+\beta^2)\tl{C}_2$ which is the direct
translation of the constraint \Ref{simple} once we have chosen the quantization
\Ref{quant3}.

\ni
Then, we have

\begin{equation}
[A^{IJ}_a,\Pi^b_{KL}]=i\hbar\delta_a^b
\left(
\gamma\delta_{KL}^{IJ}+\f{1}{2}\epsilon^{IJ}{}_{KL}
\right)
\label{comm}
\end{equation}

\ni
Let's introduce another field $E$:

\begin{equation}
E^c_{IJ}=\epsilon^{abc}
\f{1}{2}\epsilon_{IJKL}e^K_ae^L_b
\end{equation}
Then we have the following commutation relation

\begin{equation}
[A^{IJ}_a,E^b_{KL}]=i\hbar\gamma\delta_a^b\delta_{KL}^{IJ}
\end{equation}


\ni
This commutator
corresponds to a classical Poisson bracket

\begin{equation}
\{A^{IJ}_a(x),E^b_{KL}(y)\}=\gamma\delta_a^b\delta_{KL}^{IJ}
\delta(x-y)
\label{poisson}
\end{equation}

\ni
This is to be compared with the Poisson bracket of the canonical approach
(loop quantum gravity) which leads directly to a factor $\gamma$
in the spectrum of the area \cite{primer}:

\beq
\{A^i_a(x)=\Gamma^i_a+\gamma K^i_a,E^b_j(y)\}=
\gamma \delta^i_j\delta_a^b\delta(x-y)
\eeq

\ni
The two expressions share the same factor $\gamma$. What we have done
is transfer the transformation done on the spin-connection $\Gamma$
to a transformation on the tetrads. This is why, in a way, \Ref{quant3}
seems to be a natural choice of operator for $\Pi_{IJ}^a$.
Nevertheless, the Immirzi parameter is here introduced
in the area spectrum in an artificial way by a normalisation. 
Thus, it would be interesting to derive the Poisson bracket \Ref{poisson}
from a direct treatment of the simplicity constraint at the classical
level through a study of the corresponding second-class constraints
in order to understand the choice of quantization for the $\Pi$ field.




\mn
The Lorentzian case works mostly the same way as the Euclidian case.
We choose:

\begin{equation}
\Pi^a_{IJ}\rightarrow
-i\hbar \left(
\alpha\f{\delta}{\delta A^{IJ}_a}-\beta\f{\delta}{\delta *A^{IJ}_a}
\right)
\approx \hbar \tl{J}_{IJ}
\label{quantlorentz}
\end{equation}

\ni
Then, we get:

$$
\left\{\begin{array}{ccc}
{\cal B}^2&=&(\beta^2-\alpha^2){\cal A}^2\\
{\cal C}^2&=&-2\alpha\beta{\cal A}^2
\end{array}\right.
\Rightarrow
2a_2\tl{C}_2-a_1\tl{C}_1=0
$$

\ni
We can take $\alpha=1$ and normalise \Ref{quantlorentz} by
$\gamma=1/\beta$.
Then we get the same
factor $\gamma$ for the area spectrum as in the Euclidian case:

\beq
{\cal A}=\gamma \sqrt{C_1}=
\gamma\sqrt{j^2-1} \tr{ or } i\gamma\sqrt{\rho^2+1}
\eeq

\ni
This spectrum raises the question of how to interpret the imaginary
eigenvalues of the area. Should they be interpreted as time-like
components or should they be ignored? Or does it tell us we can't
build the area that way since it isn't a hermitian operator?

\mn
To conclude, the Immirzi parameter doesn't change the
representations used in the spin foam model, so it doesn't
affect the geometry of the space-time.
There is also no reason to introduce it in
a area spectrum in the covariant framework since
the simplicity constraint is independent of rescaling.
Nevertheless, in a canonical framework,
the choice of quantization \Ref{quant3}
leads to the same factor equal to the Immirzi parameter
in the spectrum of the area as in loop quantum gravity.
We should check if the choice \Ref{quant3} and the resulting
Poisson bracket \Ref{poisson} can be recovered by
perfoming a hamiltonian treatment directly on the classical BF action.
To understand what happens,
we should also investigate in more details the link between
the {\it spin foam quantization} (through the Barrett-Crane procedure
or the discretisation of BF theory) and the {\it canonical approach}
(loop quantum gravity).
We could then possibly apply the same procedure to higher-dimensional
constrained BF theory as proposed in \cite{laurent2} in order to
quantize $D\ge 4$ theories of gravity without breaking
explicitely the Lorentz invariance.

\section*{acknowledgments}

I am especially grateful to Daniele Oriti for many useful discussions.
I am also thankful to Lee Smolin, Carlo Rovelli and Alejandro Perez
for their useful comments and suggestions.

\appendix

\section*{Equivalence of the simplicity conditions}

\ni
We would like to prove the equivalence of the two simplicity
conditions \Ref{simple} and \Ref{simple2}. We are going to follow
the proof given in \cite{so4} for the case $a_1=0$ and we will restrict
ourselves to the Euclidian case. This proof can be adapted to the
Lorentzian case provided we change some signs (due to $*^2=-1$).

\mn
First, we define:

\beq
e=\f{1}{4!}\epsilon_{IJKL}B^{IJ}\w B^{KL} 
\eeq

\ni
Then supposing $a_2=0$,
the constraints \Ref{whole} and \Ref{simple} on the $B$
field can be written as:

\beq
\epsilon^{abcd}B^{IJ}_{ab}B^{KL}_{cd}=
e\left(
\epsilon^{IJKL}+\f{a_1}{a_2}\eta^{[IK}\eta^{J]L}
\right)
=e\Omega^{IJKL}
\label{app1}
\eeq

\ni
By raising and lowering indices, we can rewrite \Ref{app1} 
under a more convenient form in which we consider
all the variables as $6\times 6$ matrices
(on the antisymmetric couples $[IJ]$):

\beq
B_{IJ}^{ab}\epsilon_{ab}^{cd}B_{cd}^{KL}=e\Omega_{IJ}^{KL}
\label{app2}
\eeq

\ni
Then $\Omega_{IJ}^{KL}$ is invertible when

$$
\left(\f{a_1}{2a_2}\right)^2\ne 1
$$

\ni
In that case, assuming that
$e\ne0$ (the B field is non-degenerate),
we can define

\beq
\Sigma^{ab}_{IJ}=
\f{1}{e}\epsilon^{abcd}
\f{2}{\left(\f{a_1}{a_2}\right)^2-4}
\left(
-\epsilon_{IJKL}+\f{a_1}{2a_2}\eta_{[IK}\eta_{J]L}
\right)
B^{KL}_{cd}
\label{sigma}
\eeq

\ni
and rewrite \Ref{app2} under the simple form:

\beq
\Sigma^{cd}_{IJ}B^{KL}_{cd}=\delta^{KL}_{IJ}
\label{app3}
\eeq

\ni
\Ref{app3} means that 
$\Sigma_{IJ}^{cd}$ and $B_{cd}^{KL}$
are invertible and
inverse of each other. This means that \Ref{app3}
is equivalent to:

\beq
B_{cd}^{IJ}\Sigma_{IJ}^{ab}=\delta_{cd}^{ab}
\label{app4}
\eeq

\ni
After writing explicitely this new equation
and passing the tensor $\epsilon^{abcd}$ from the LHS
to the RHS, we find:

\beq
2\left(
-\epsilon_{IJMN}+\f{a_1}{2a_2}\eta_{[IM}\eta_{J]N}
\right)
B^{MN}_{cd}B^{IJ}_{ab}=
\epsilon_{abcd}
e\left(\left(\f{a_1}{a_2}\right)^2-4\right)
\eeq

\ni
and we recover the simplicity condition
\Ref{simple2} in the case $(ab)=(cd)$.

\mn
Let's note that in the Lorentzian case,
the condition on $a_1,a_2$ reads

$$
\left(\f{a_1}{2a_2}\right)^2\ne -1
$$

\ni
and is automatically satisfied if we keep real variables.

\end{document}